\documentstyle[sprocl,epsfig]{article}

\def\NPB{{\em Nucl. Phys.} B}
\def\PLB{{\em Phys. Lett.}  B}

\def\be{\begin{equation}}
\def\ee{\end{equation}}
\def\bea{\begin{eqnarray}}
\def\eea{\end{eqnarray}}

\newcommand{\eq}{\begin{equation}}
\newcommand{\en}{\end{equation}}

\newcommand{\bdm}{\begin{displaymath}}
\newcommand{\edm}{\end{displaymath}}

\newcommand{\ba}{\begin{array}}
\newcommand{\ea}{\end{array}}

\newcommand{\ZZ}{\hbox{{\rm Z{\hbox to 3pt{\hss\rm Z}}}}}
\newcommand{\Br}{\langle}
\newcommand{\kt}{\rangle}

\newcommand{\tq}{\frac34}

\begin{document}

\title{PROBING THE INTERIOR OF THE COLOUR FLUX TUBE}

\author{F. GLIOZZI}
\address{Dipartimento di Fisica Teorica, Universit\`a 
di Torino, \\ 
        via P. Giuria 1, 10125 Torino, Italy}
       
\maketitle\abstracts{In the dual superconductivity description of quark
confinement the core of the flux tube connecting a quark pair belongs to a
deconfined, hot phase. This can be checked in numerical experiments 
on $3D$ $Z_2$ gauge model. The Svetitsky-Yaffe conjecture provides 
analytic expressions for the distribution of the flux density around quark
sources at  critical temperature. }

\section{Introduction}

The internal structure of the colour flux tube (CFT) joining a quark pair in 
the confining phase of any gauge model provides an
important test of the dual superconductivity (DS) conjecture \cite{tmp}, 
because it should show, as the dual of an Abrikosov vortex, 
a  core of normal, hot vacuum as contrasted with the surrounding 
medium, which is in the dual superconducting phase.  A general way to study 
the internal structure of the flux tube is to test it with suitable gauge 
invariant probes. More specifically, the vacuum state of a lattice gauge model 
is modified by the insertion in the action of a quark source (for instance a 
Wilson loop). In this modified vacuum (called W-vacuum) one can evaluate the
expectation value of various probes as a function of their position with respect
the quark sources. Some general results of such an analysis has been already
reported in Ref.2. Here I will describe some new results which are
specific of the $3D$ $\ZZ_2$ gauge model.

\section {The Disorder Parameter around Quark Sources}
The location of the core of the CFT is  given in  DS conjecture by the 
vanishing of the disorder parameter 
$\Br\Phi_M(x)\kt$, where $\Phi_M$ is some effective magnetic Higgs field. 
In a pure gauge theory, the formulation of this property poses some problems, 
because in general no local, gauge invariant, disorder field $\Phi_M(x)$ is 
known. In the special case of $3D$ $\ZZ_2$ gauge model there is an exact 
duality, namely the Kramers-Wannier transformation, which maps the gauge 
theory in the Ising model. The spontaneous magnetization $\mu=\Br\sigma\kt$ is 
precisely the wanted disorder parameter: it vanishes in the deconfined phase, 
while it is different from zero  in the confining phase.
As an example, in Fig.1  the spontaneous magnetization in a W-vacuum generated 
by a pair of parallel Polyakov loops is reported.
One can clearly see the formation of a flux tube with a core where the disorder
parameter vanishes, as required by the DS conjecture.
\vskip -2.3cm
\hskip 1cm\epsfig{file=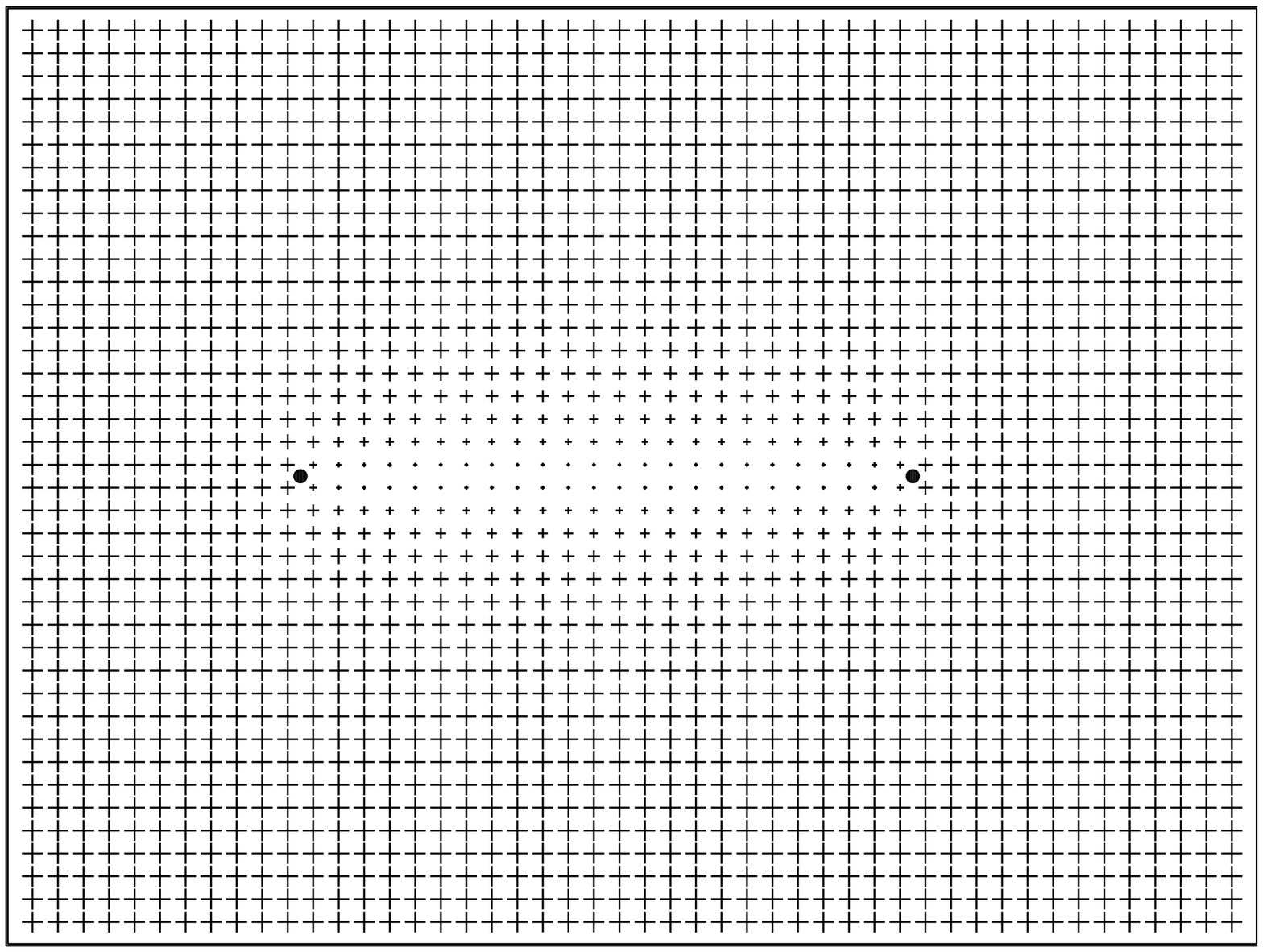,height=10.5cm}
\vskip -1.9cm
{\hskip1.4cm\footnotesize Figure 1. Spontaneous magnetization around a 
quark pair.}
\vskip0.2cm
\vskip -2.3 cm
\hskip 1cm\epsfig{file=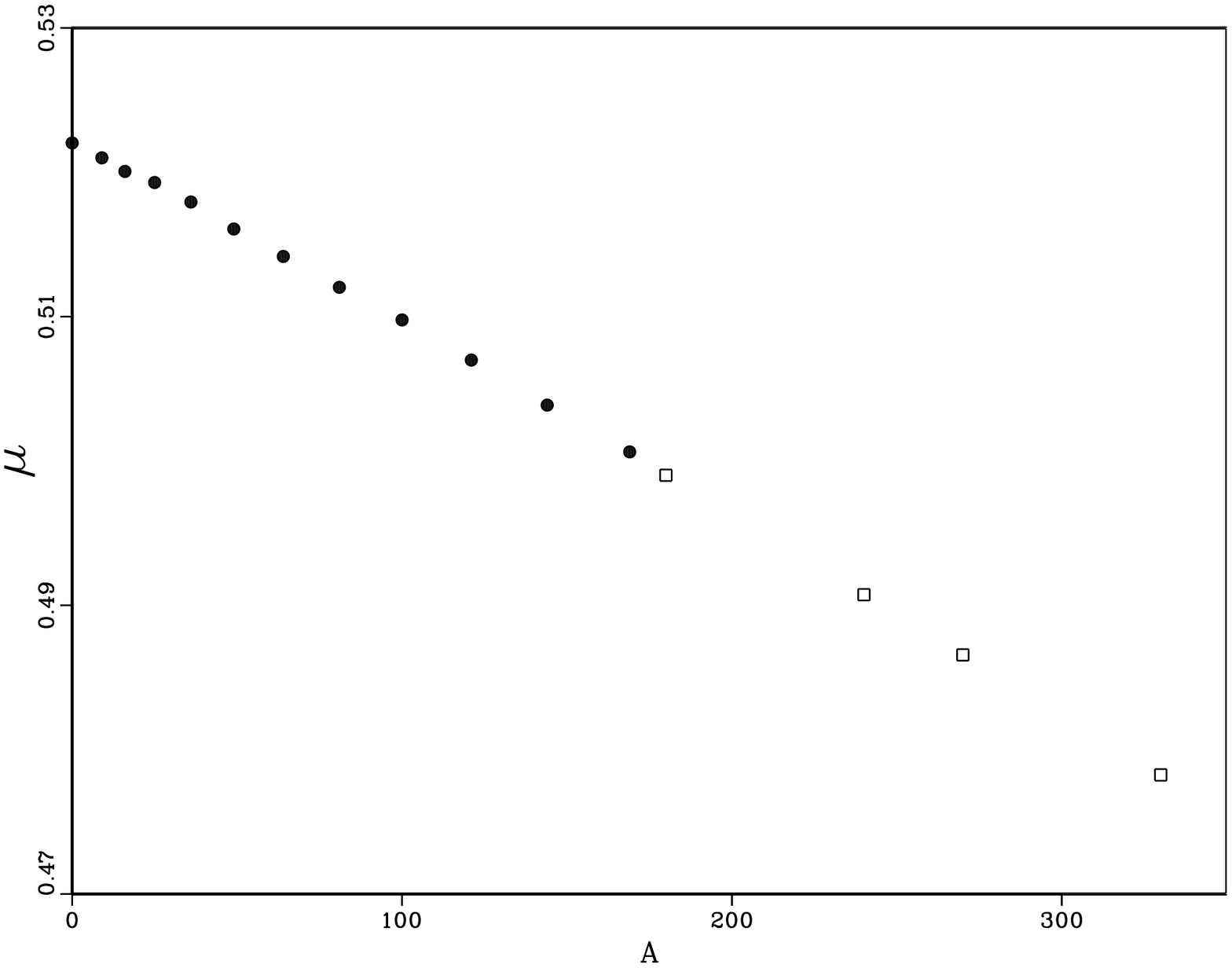,height=10.5cm}
\vskip -1.6cm
\noindent
{\footnotesize Figure 2.  Total  magnetization as a function of the loop area.
The black dots are square Wilson loops, the open symbols are pairs of Polyakov
loops.}
\vskip0.2cm

The total thickness of the flux tube is the sum of two different contributions:
one is due to quantum fluctuations of string-like modes of the CFT, which
produce an effective squared width growing logarithmically with the interquark
distance \cite{lmw,width}; the other is the intrinsic thickness of the flux
tube, which according to the DS conjecture is non-vanishing. 

The total magnetization of the W-vacuum provides us with a method to evaluate 
such an intrinsic thickness: describing  the CFT approximately as a cylinder of
vanishing magnetization  immersed in a mean of 
magnetization $\mu\not=0$  we get that the total
magnetization of the W-vacuum in a finite volume decreases linearly with the
volume $V$ spanned by the CFT as shown in Fig.2 , with $V= A L_c$, where 
$L_c$ is the intrinsic thickness of the tube and $ A$ is the  area 
of the minimal surface bounded by the Wilson loop (black dots) or by a 
Polyakov pair (open squares).
The slope of such a linear behaviour yields an intrinsic thickness 
$L_c\sqrt{\sigma}=0.98(2)$  ($\sigma$ is the string tension) in  reasonable 
agreement with the theoretical value of $\sqrt{\pi/3}$ suggested by a 
conformal field theory argument \cite{cg}.

\section{The Colour Flux Tube at Criticality}

 According to the widely tested Svetitsky-Yaffe conjecture \cite{sy},  any 
gauge theory in $d+1$ dimensions with a continuous deconfining transition 
belongs to the same universality class of a $d$-dimensional $C(G)$-symmetric 
spin model, where $C(G)$ is the center of the gauge group. 
It follows that at the critical 
point all the critical indices describing the two transitions and all the 
adimensional ratios of correlation functions of corresponding observables 
in the two theories should coincide. 

In particular, since the order parameter the gauge theory is 
mapped into the corresponding one of the spin model, the correlation functions  
among  Polyakov loops should be proportional to the corresponding correlators 
of spin operators:
\eq
\Br P_1\dots P_{n}\kt_{T=T_c}\propto \Br \sigma_1\dots \sigma_{n}\kt~~.
\en
The crucial point is that for $d=2$ the form of these universal functions is 
exactly known. Then one can use these analytic results to get useful 
informations on the internal structure\cite{gv} of the colour flux tube at 
$T=T_c$. For instance, the correlator 
\eq
\Br P_1\dots P_{n+2}\kt=\Br P(x_1,y _1)\dots P(x,y)P(x+\epsilon,y)\kt~~,
\en
thought as a function of the spatial coordinates $x,y$  of the last two
Polyakov loops (used as probes), describes, when $\epsilon$ is chosen small 
with respect to the other distances entering into the game, the distribution 
of the flux around  $n$ Polyakov loops with spatial coordinates 
$x_i,y_i$ $(i=1,\dots n)$.
In Fig.3 the contour lines of the flux distribution 
$\rho(x,y)=\Br P_1\dots P_6\kt/\Br P_1\dots P_4\kt-\Br P_5 P_6\kt$
in a critical gauge system with $C(G)=\ZZ_2$ are reported. The Polyakov lines 
are located at the corners of a rectangle $d\times r$ with $d>r$. Denoting by 
$r_i$ the distance of the probe $(P_5 P_6)$ from the source $P_i$ one has
simply
\eq
\rho(x,y)\propto\epsilon^{\tq}\left\{\frac{rd}{\sqrt{r^2+d^2}}
\frac{\sum_i r_i^2}{\prod_ir_i}+O(\epsilon)\right\}~~~.
\en
One clearly sees the formation of two flux tubes connecting the two pairs of 
nearest sources.
Comparison with the distribution obtained by the sum of the fluxes generated by
two non-interacting ($i.e.\;d=\infty$) flux tubes (dotted contours) indicates an 
attractive interaction between them, as expected.
\vskip -2.3 cm
\hskip1cm\epsfig{file=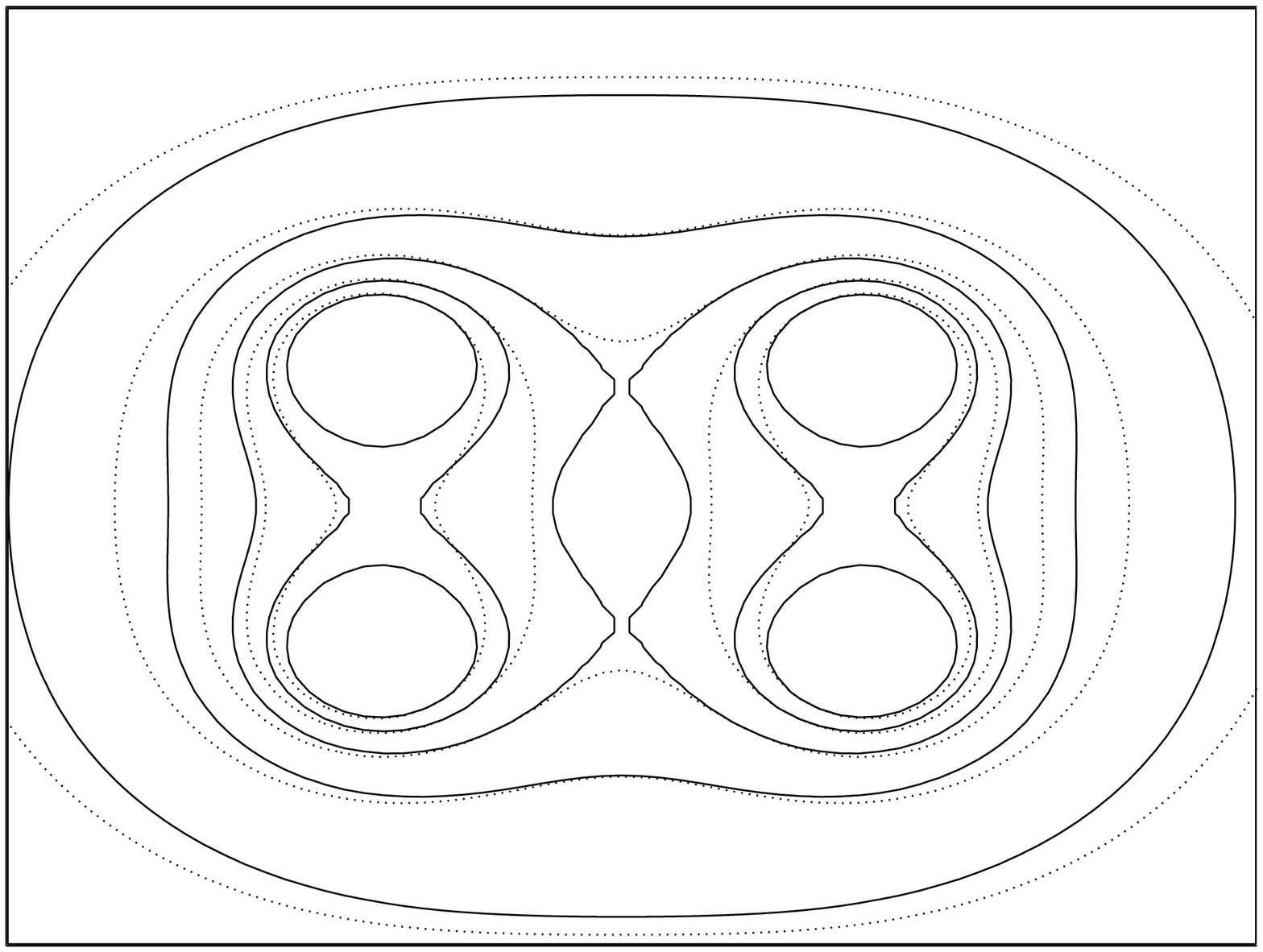,height=10.5cm}
\vskip -1.6cm
\noindent
{\footnotesize Figure 3. Contours of flux density around two pairs of 
parallel Polyakov loops at criticality. The dot\-ted lines correspond to the 
contours in the non-interacting case.}

\end{document}